\documentclass[12pt]{iopart} 
\usepackage{epsfig,psfrag}
\usepackage{amsmath,amsfonts,amssymb}
\usepackage{bbm}
\usepackage{mathrsfs}

\newcommand{\Ecal}{\mathcal E}
\newcommand{\Mcal}{\mathcal M}

\newcommand{\xbf}{\mathbf {x}}

\newcommand{\pbf}{\mathbf {p}}
\newcommand{\qbf}{\mathbf {q}}

\newcommand{\sbf}{\mathbf {s}}
\newcommand{\ubf}{\mathbf {u}}
\newcommand{\nbf}{\mathbf {n}}

\newcommand{\pd}{\partial}
\newcommand{\til}[1]{\tilde{#1}}

\begin{document} 
\title{
Photon density of states for deformed surfaces
}
\author{T Emig$^1$}
\address{$^1$ Institut f\"ur Theoretische Physik, Universit\"at zu
K\"oln, Z\"ulpicher Stra\ss e 77,\\ D-50937 K\"oln, Germany}
\ead{te@thp.uni-koeln.de} 
\begin{abstract} 
  A new approach to the Helmholtz spectrum for arbitrarily shaped
  boundaries and a rather general class of boundary conditions is
  introduced. We derive the boundary induced change of the density of
  states in terms of the free Green's function from which we obtain
  both perturbative and non-perturbative results for the Casimir
  interaction between deformed surfaces. As an example, we compute the
  lateral electrodynamic Casimir force between two corrugated surfaces
  over a wide parameter range. Universal behavior, fixed only by the
  largest wavelength component of the surface shape, is identified at
  large surface separations. This complements known short distance
  expansions which are also reproduced.
\end{abstract} 
\submitto{\JPA} 

\section{Introduction}

Casimir interactions can be viewed as a consequence of a change in the
photon spectrum under a relative displacement of the interacting
objects. For two parallel ideal metallic plates the change in the
photon density of states (DOS) with the plate distance $H$ is
well-known \cite{Balian+78}. It is peaked close to the characteristic
frequency $\omega\sim c/H$ which provides the dominant contribution to
the interaction. However, even for simple geometries like a plate and
a sphere no exact expression for the change in the DOS is
available. The reason for that is the difficulty to compute the
distribution of eigenfrequencies of the Helmholtz wave equation in
arbitrary geometries. This is inherently linked to the problem of
chaotic (quantum) billiards \cite{chaos-book}.  For Casimir
interactions the natural question arises to what extent the force
characterizes the shape of the interacting objects and vice versa.
There is a long debate about the existence of a repulsive Casimir
interaction between ideal metals which was mainly initiated by the
positivity of the Casimir energy of a single infinitesimal thin sphere
of ideal metal \cite{Boyer}. So far, no example of a repulsive force
between two separated ideal metallic bodies has been found. 

During the last decades a number of approximative methods have been
developed to determine the Casimir interaction in more complicated
geometries. These include the proximity and pairwise additivity
approximations \cite{Bordag+01}, semiclassical approaches
\cite{Schaden+98}, multiple scattering expansions \cite{Balian+78}
and, more recently, a simple to implement optical approach
\cite{Jaffe+04} which is however limited to short distances. Despite
the usefulness of these approaches in certain limits, reliable results
for the entire range of energies (or distances) are not known in
general. Here I shall present a formula for the density of states in
terms of the free space Green's function which is evaluated at the
(ideal metallic) surfaces only.  This formula will be used to compute
the DOS for two deformed surfaces both perturbatively in the
deformations and also exactly with the aid of a previously developed
numerical method for periodic surfaces \cite{Emig+04}. As an
application, we study the lateral force between periodic surfaces.  It
is found that the surface's shape can be deduced from the force at
short distances whereas at large separations universality prevails.

\section{Trace formula for the photon spectrum}

For geometries with a translational invariant direction the
electromagnetic field splits into transverse electric (TE) and
magnetic (TM) modes. Both modes are described by a scalar field which
fulfills either Dirichlet or Neumann boundary conditions on the ideal
metal surfaces. Hence, we consider in the following the scalar field
Helmholtz equation
\begin{equation}
(\nabla^2 + k^2) \phi({\bf x}) = 0 \, .
\end{equation}
The relevant quantity for the Casimir interaction is the {\it change}
$\delta\rho(k)=\rho(k)-\rho_\infty(k)$ in the DOS caused by moving the
surfaces from infinity to the finite distance for which one 
is interested in
the Casimir energy. Hence $\rho_\infty(k)$ is the DOS for
infinitely separated surfaces. The relevant finite part of the
zero-point energy is then given by the sum of $k
\delta\rho(k)$ over $k$. It is more convenient to work in Euclidean space
by performing a Wick rotation to imaginary frequencies. With
$\delta\tilde\rho(q_0)=-\delta\rho(iq_0)$ the Casimir energy 
becomes
\begin{equation}\label{eq:general-energy}
{\cal E}\:=\:\frac{\hbar c}{2}\int_0^\infty dq_0\,q_0\,
\delta\tilde\rho(q_0) \, .
\end{equation}
In order to derive the DOS we start with the 4-dimensional Euclidean
action
\begin{equation}
S=\frac{1}{2} \int d^4 X (\nabla \phi)^2 \, .
\end{equation}
Using a path-integral formulation with delta-functions enforcing the
boundary conditions on the surfaces $S_\alpha$ \cite{Li+92}, one
obtains the two-point correlation function in the presence of
boundaries \cite{Emig+05}. With
 the free
space Green's function $G_0(\xbf,\xbf';q_0)=e^{-q_0|\xbf-\xbf'|}/(4\pi|\xbf-\xbf'|)$, the modified correlations
$G(\xbf,\xbf';q_0)=\langle \phi_{q_0}(\xbf)\phi_{-q_0}(\xbf')\rangle$
in the presence of boundaries is given by
\begin{eqnarray}
\label{eq:boundary-green}
G(\xbf,\xbf';\!q_0) &-& G_0(\xbf,\xbf';\!q_0)
 = -\sum_{\alpha\beta} 
\int d\ubf \, d\ubf' \,
G_0(\xbf,\sbf_\alpha(\ubf);\!q_0) \nonumber \\
&\times&\Mcal^{-1}_{\alpha\beta}(\ubf,\ubf';\!q_0)\, 
G_0(\xbf',\sbf_\beta(\ubf');\!q_0) \, ,
\end{eqnarray}
where $\Mcal^{-1}_{\alpha\beta}(\ubf,\ubf';\!q_0)$ is the functional
inverse of $\Mcal_{\alpha\beta}(\ubf,\ubf';q_0)=G_0(\sbf_\alpha(\ubf),
\sbf_\beta(\ubf');q_0)$ with the 3D vectors $\sbf_\alpha(\ubf)$
denoting the positions of surface $S_\alpha$ as a function of the 2D
coordinate $\ubf$.  This result applies to Dirichlet boundary
conditions. For the Neumann case $G_0$ has to be replaced by
$\pd_{\nbf_\alpha(\ubf)}G_0(\xbf,\sbf_\alpha(\ubf);\!q_0)$ on the
right-hand side of Eq.~\eqref{eq:boundary-green} and
$\Mcal_{\alpha\beta}(\ubf,\ubf';q_0)=\pd_{\nbf_\alpha(\ubf)}
\pd_{\nbf_\beta(\ubf')} G_0(\sbf_\alpha(\ubf),\sbf_\beta(\ubf');q_0)$
with $\pd_{\nbf_\alpha}$ the normal derivative at the surface
$S_\alpha$. Next we make use of the property that the trace of
Green's function yields the DOS. In the Euclidean formulation
this relation reads 
\begin{equation}
\rho(iq_0)= \frac{2q_0}{\pi} \int d^3 \xbf \, G(\xbf,\xbf;q_0) \, .
\end{equation}
Here the spatial integration extends over all inner and other regions
of the surfaces. Since Eq.~\eqref{eq:boundary-green} holds in any
spatial region, we obtain the change in the DOS by integrating the
right-hand side of that equation with $\xbf=\xbf'$ running over the
entire 3D space. The integration is easy to perform in momentum space
since $\xbf$ and $\xbf'$ occur as arguments of the free Green's function in
Eq.~\eqref{eq:boundary-green}. Using for the Fourier transformed
propagator $\partial/\partial q_0 (q_0^2+q^2)^{-1} = -2q_0
(q_0^2+q^2)^{-2}$, one obtains after $\xbf$-integration both for $\Mcal$ and
$\Mcal_\infty$ the final trace formula
\begin{equation}
\label{eq:trace-formula}
\delta\tilde\rho(q_0)=-\frac{1}{\pi}\frac{\partial}{\partial q_0}
\textrm{Tr}\ln\left({\cal M}{\cal M}^{-1}_\infty\right) \, .
\end{equation}
The trace runs over the 2D coordinates $\ubf$ and the surface indices
$\alpha$.  An advantage of this formula is its independence of surface
self-energies which diverge in the limit of ideal metals but are
irrelevant for the interaction.  It is important to note that this
formula can be applied in any spatial dimension and also to other
types of boundary conditions which can be implemented via
delta-functions in the path integral \cite{Buescher+04}.

\section{Density of states}

In this section we will apply the trace formula of
Eq.~\eqref{eq:trace-formula} to two deformed but on average parallel
plates. In the first part we study (in amplitude and curvature) small
deformations of general shape (corresponding to the large distance
limit) by perturbation theory.  Non-perturbative numerical results for
the DOS of periodically corrugated surfaces with edges at closer
distances are presented in the second part.

\subsection{Perturbation theory for arbitrarily deformed surfaces}

\begin{figure}[t]
\begin{center}
\includegraphics[width=0.7\linewidth]{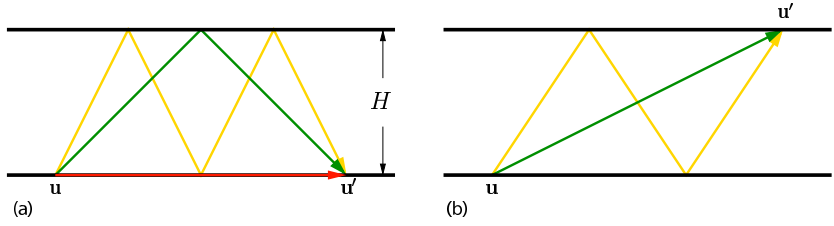}
\caption{\label{fig:fig1}Graphical representation of the contributions
  to the kernels $K_\|$ (a) and $K_\times$ (b),
  cf.~Eq.~\eqref{eq:kernels}.}
\end{center}
\end{figure}

Let us consider two parallel surfaces with small deformations which
are parametrized by $\sbf_1(\ubf)=(\ubf,h_1(\ubf))$ and
$\sbf_2(\ubf)=(\ubf,H+h_2(\ubf))$ where the deformation fields
$h_\alpha$ are small compared to the mean distance $H$. We are
interested in the deformation induced modification of the flat surface
DOS
\begin{equation}
\label{eq:dos_flat}
\delta\tilde\rho_0(q_0)=\frac{q_0}{2} \log \left(
1-e^{-2q_0 H}\right) A \, ,
\end{equation}
where $A$ is the surface area. Expanding the kernel $\Mcal = \Mcal_0
+\delta\Mcal$ with $\Mcal_0$ the kernel for flat surfaces and
$\delta\Mcal$ the correction due to the deformations, one finds
for the change of the DOS
\begin{equation}
\delta \tilde \rho(q_0) =  
-\frac{1}{\pi} \text{Tr} \left[
\log \Mcal_0 + \log (1+\Mcal_0^{-1} \delta\Mcal)
\right] - ``H\to\infty'' \, ,
\end{equation}
where the contribution for $H\to\infty$ has to be subtracted. 
Expanding the logarithm to second order in the deformations
$h_\alpha(\ubf)$, the change of the DOS can be written as
\begin{eqnarray}
\label{eq:dos-pert-exp}
\delta\tilde\rho(q_0)&=& \delta\tilde\rho_0(q_0)
-\frac{1}{4\pi^2}\frac{q_0^3}{\sinh^2(q_0H)}
\int_\ubf \left[ h_1^2(\ubf)+h_2^2(\ubf) \right]\nonumber\\
&+& \int_{\ubf,\ubf'} \left\{
\frac{1}{2} \sum_\alpha K_\|(q_0;\ubf-\ubf') \left[ h_\alpha(\ubf)
-h_\alpha(\ubf')\right]^2\right. \nonumber\\
 &+& \left. \sum_{\alpha\neq\beta}
K_\times(q_0;\ubf-\ubf') h_\alpha(\ubf)h_\beta(\ubf')
\right\} \, ,
\end{eqnarray}
where $\delta\tilde\rho_0$ is given by Eq.~\eqref{eq:dos_flat}.  The
kernels of the non-local contributions can be expressed in terms of
the free Green's function. In the following we will focus due to
limitations in space on Dirichlet conditions; the Neumann case has the
form of Eq. \eqref{eq:dos-pert-exp} with however different kernels.
Defining the series
\begin{eqnarray}
\label{eq:R-series}
R_\text{even}(q_0;u)&=&\sum_{n=1}^\infty 
\frac{G_0(\sqrt{u^2+(2nH)^2};q_0)}{(2n)^2}, \nonumber \\ 
R_\text{odd}(q_0;u)&=&\sum_{n=1}^\infty 
\frac{G_0(\sqrt{u^2+((2n-1)H)^2};q_0)}{(2n-1)^2}
\end{eqnarray}
the kernels can be written as
\begin{eqnarray}
\label{eq:kernels}
K_\|(q_0;\ubf) &=& -\frac{8}{\pi} \partial_{q_0} \left[
\left( \frac{1}{u}\partial_u G_0(u;q_0) + \partial^2_H
R_\text{even}(q_0;u)\right)\partial_H^2 R_\text{even}(q_0;u)
\right] \nonumber \\
K_\times(q_0;\ubf) &=& -\frac{8}{\pi} \partial_{q_0} \left[
\partial_H^2 R_\text{odd}(q_0;u)\right]^2 \, .
\end{eqnarray}
The contributions to the kernels can be viewed as multiple scattering
paths between flat surface positions. The kernel $K_\|$ has only
contributions from an even number of paths crossing the gap,
connecting two positions on the same surface, see
Fig.~\ref{fig:fig1}(a). The first factor $\sim \partial_u G_0$ in
$K_\|$ is represented as a non-crossing paths. An odd number of
crossing paths yields the kernel $K_\times$, see
Fig.~\ref{fig:fig1}(b). Each path is associated with the free Green's
function which is reduced by a factor which is the inverse squared
number of gap crossings, cf.~Eq.~\eqref{eq:R-series}.

\subsection{Non-perturbative treatment of periodic surfaces}

The above perturbative treatment is limited to small and smooth
deformations. However, the trace formula turns out to be useful also
in the case of surfaces with edges, and at large deformation
amplitudes. Recently, a numerical tool has been developed to evaluate
the Casimir force for periodic surface deformations \cite{Emig+04}. Here
we shall demonstrate that same approach can be also applied to the
DOS. Specifically, we consider uni-axial rectangular surface profiles
shown in Fig.~\ref{fig:fig3}(a). In this section we restrict the
analysis to no lateral shift, $b=0$.  By Fourier transforming
$\Mcal_{\alpha\beta}(\ubf,\ubf';q_0)$ with respect to $\ubf$ and
$\ubf'$ one can employ the periodicity of the surface profiles (along
the $x_1$-direction) which leads to the representation
\begin{equation}\label{eq:Bloch}
\begin{split}
  &\til{\Mcal}_{\alpha\beta}\left(\pbf,\qbf;q_0\right)\:=\:
  (2\pi)^2\delta\left(p_2 + q_2 \right)\\
  &\times\sum_{m=-\infty}^\infty \delta\left(p_1+q_1+2\pi
    m/\lambda\right)\, N_m^{\alpha\beta}\left(q_1,q_2;q_0\right) \, ,
\end{split}
\end{equation} 
which defines the $2\times2$ matrices $N_m$ that can be computed
analytically and are given in \cite{Emig+04} for the
present geometry. $\til{\Mcal}$ can be transformed to block-diagonal
form so that each block couples only waves whose momenta differ from a
given Bloch momentum by integer multiples of $2\pi/\lambda$. This 
decomposition can be used to define the function
\begin{equation}
\label{eq:g-fct}
g(q_1,q_2;q_0)=\text{tr}(B^{-1}\pd_{q_0}B - B^{-1}_\infty\pd_{q_0}B_\infty)
\end{equation}
where the trace runs over all discrete indices of the matrix
$B_{kl}^{\alpha\beta}(q_1,q_2;q_0)=N_{k-l}^{\alpha\beta}(q_1+2\pi
l/\lambda,q_2;q_0)$, and $B_\infty$ is the analog of $\Mcal_\infty$,
i.e., $B$ for $H\to\infty$. For practical computations we take $k$,
$l=-M,\ldots,M$ with an integer cutoff $M$ which should be taken to
infinity. The change of the DOS can then be expressed as
\begin{equation}
\delta\tilde\rho(q_0)=-\frac{A}{4\pi^3}\int_{-\infty}^\infty
dq_2 \int_0^{2\pi/\lambda} dq_1 \, g(q_1,q_2;q_0) \, .
\end{equation}
The following results for {\it electromagnetic} fluctuations are based
on the application of this formula to Dirichlet (TM modes) and Neumann
(TE modes) boundary conditions where the inversion of $B$, the trace
in Eq.~\eqref{eq:g-fct} and the integration are performed numerically.

\begin{figure}[h]
\begin{center}
\includegraphics[width=0.45\linewidth]{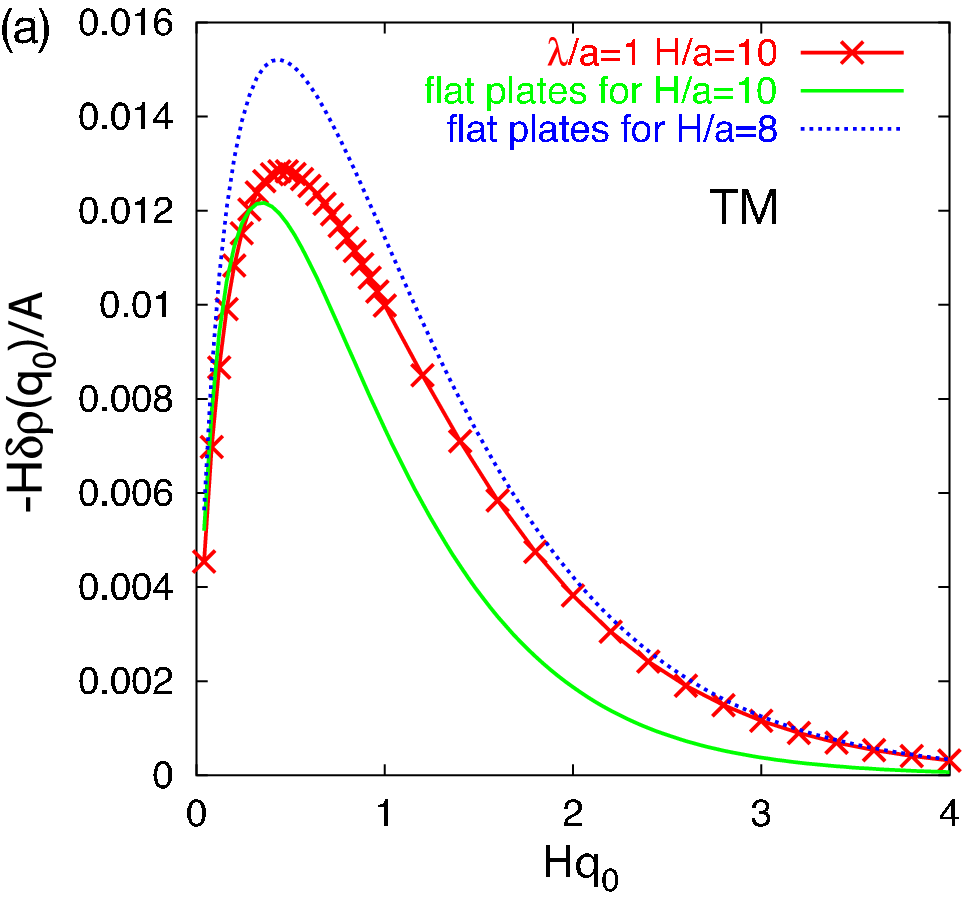}
\includegraphics[width=0.45\linewidth]{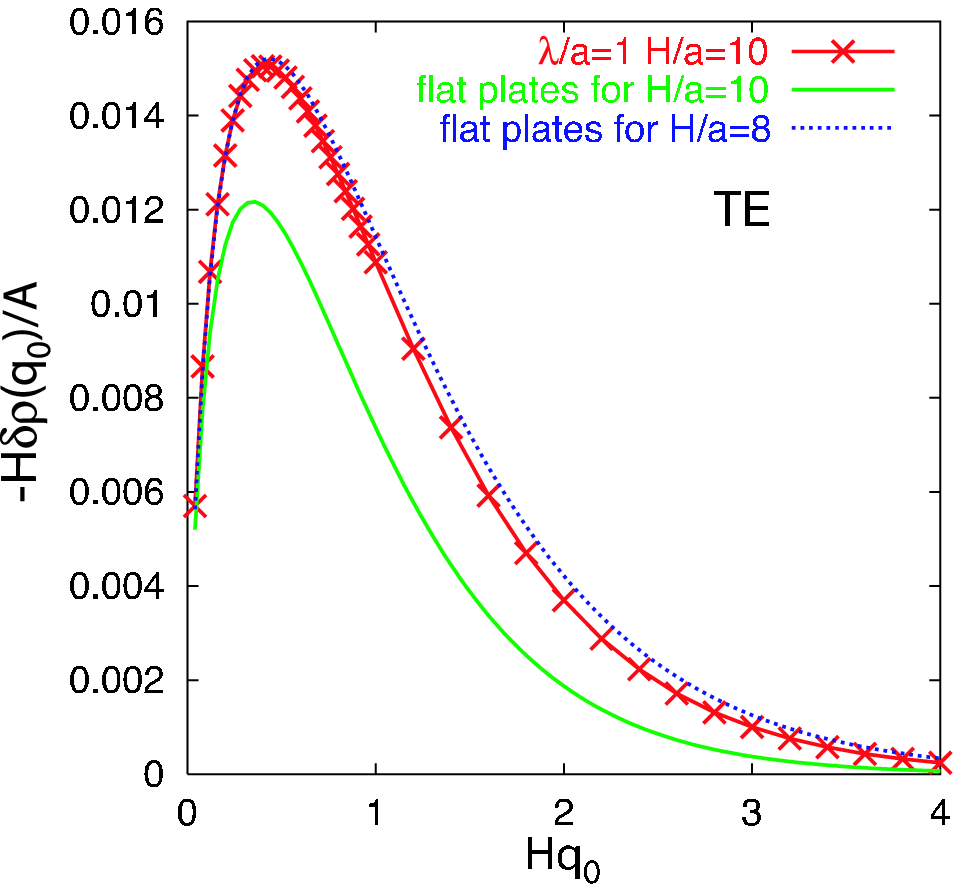}
\includegraphics[width=0.45\linewidth]{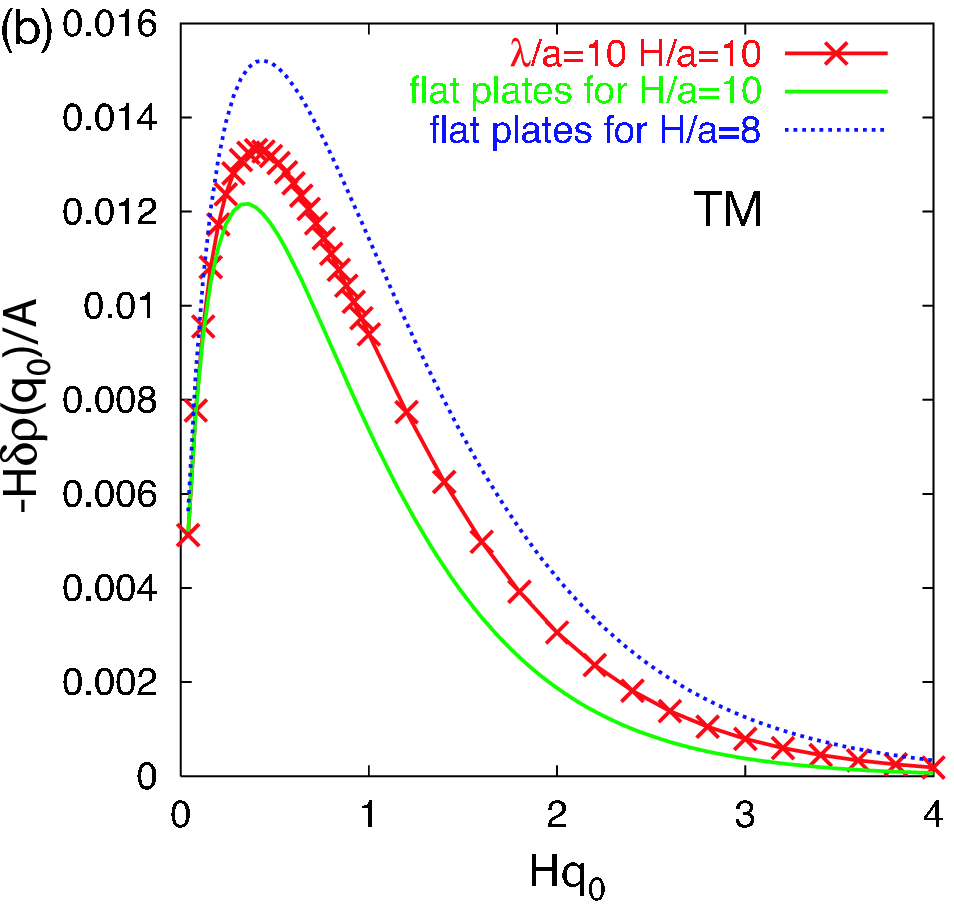}
\includegraphics[width=0.45\linewidth]{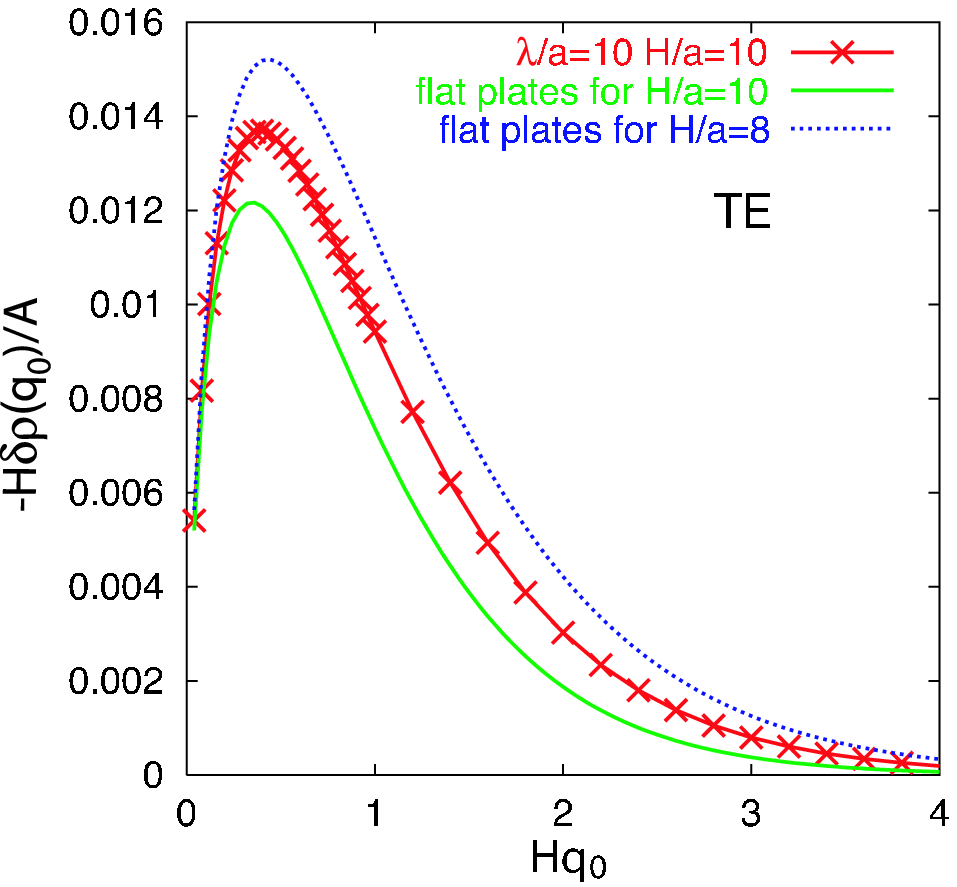}
\includegraphics[width=0.45\linewidth]{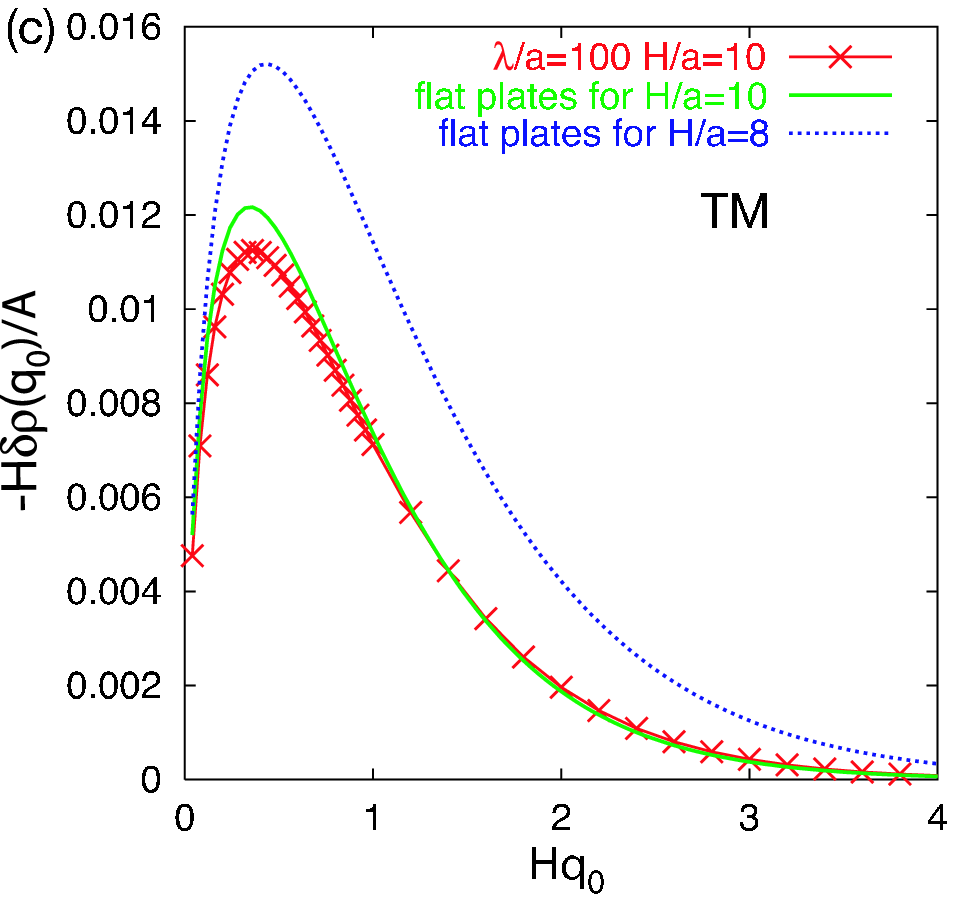}
\includegraphics[width=0.45\linewidth]{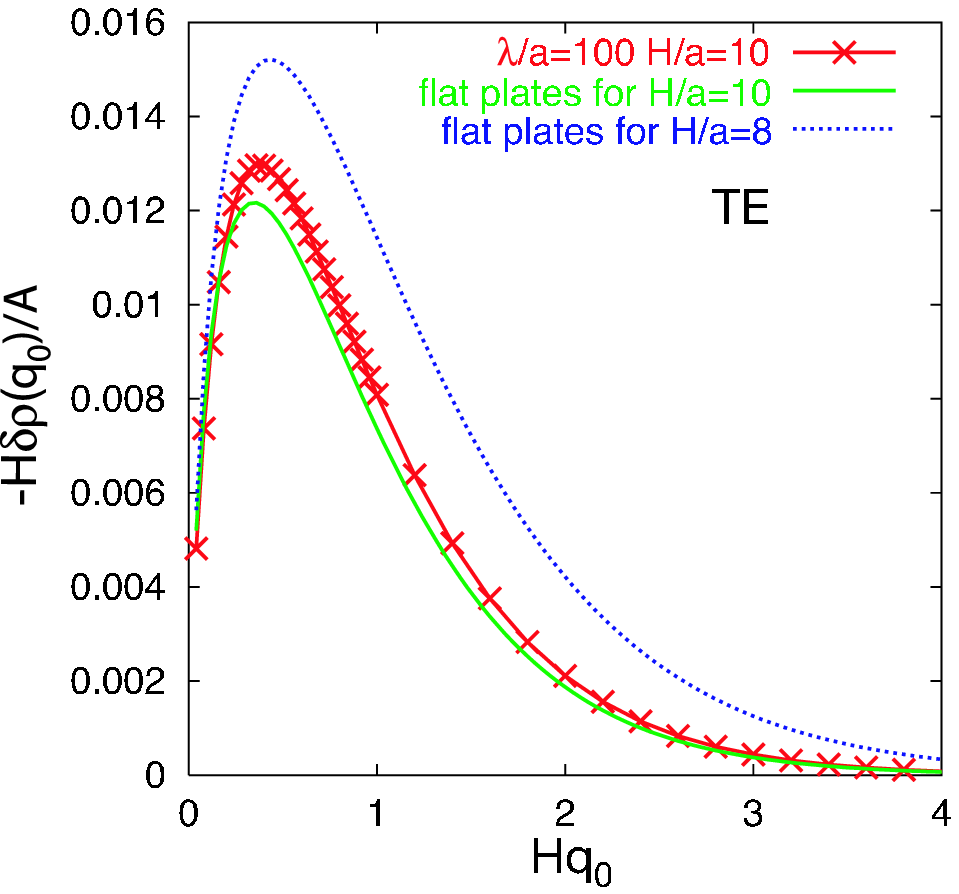}
\caption{\label{fig:fig2}Change of the density of states $\delta
  \rho(q_0)$ per unit area for TM and TE modes with parameters $H=10a$
  and $\lambda/a=1$ (a), $10$ (b), $100$ (c). The results were
  obtained with a cutoff of $M=13$. (For scaling of the ($Hq_0$)-axis,
  $H=10a$.) }
\end{center}
\end{figure}

Fig.~\ref{fig:fig2} displays the change of the DOS separately for TM
and TE modes at the fixed distance $H=10a$ and $\lambda$ ranging from
$a$ to $100a$. Plotted is also the analytic expression for {\it flat}
plates at the same mean distance and at the reduced distance $H-2a$. An
expected feature of these results is that the dominant
contribution to the change of the DOS comes from frequencies $q_0\sim
H^{-1}$. For small modulation lengths $\lambda\ll H$ one can expect
that for $q_0\lesssim \lambda^{-1}$ the modes cannot probe the narrow
valleys of the surface, and the DOS should be well described by the
result for flat plates at a reduced distance $H-2a$. While this is
fully confirmed for TE modes, the magnetic modes show an unexplained
deviation at sufficiently small $q_0$, see Fig.~\ref{fig:fig3}(a).
For $\lambda\sim H$, Fig.~\ref{fig:fig3}(b), no significant
differences between the two types of modes are observed, and the
change of the DOS is in between the result for flat surfaces at
distances $H$ and $H-2a$, respectively. For large $\lambda \gg H$,
Fig.~\ref{fig:fig3}(c), there is a low density of edges, and one can
expect the DOS for flat plates at distance $H$ to be a good
approximation.  Such behavior is indeed observed with a small
decrease for TM and a small increase for TE modes close to the
peak.  

\section{Universality of the lateral Casimir force}

\begin{figure}[t]
\begin{center}
\includegraphics[width=0.36\linewidth]{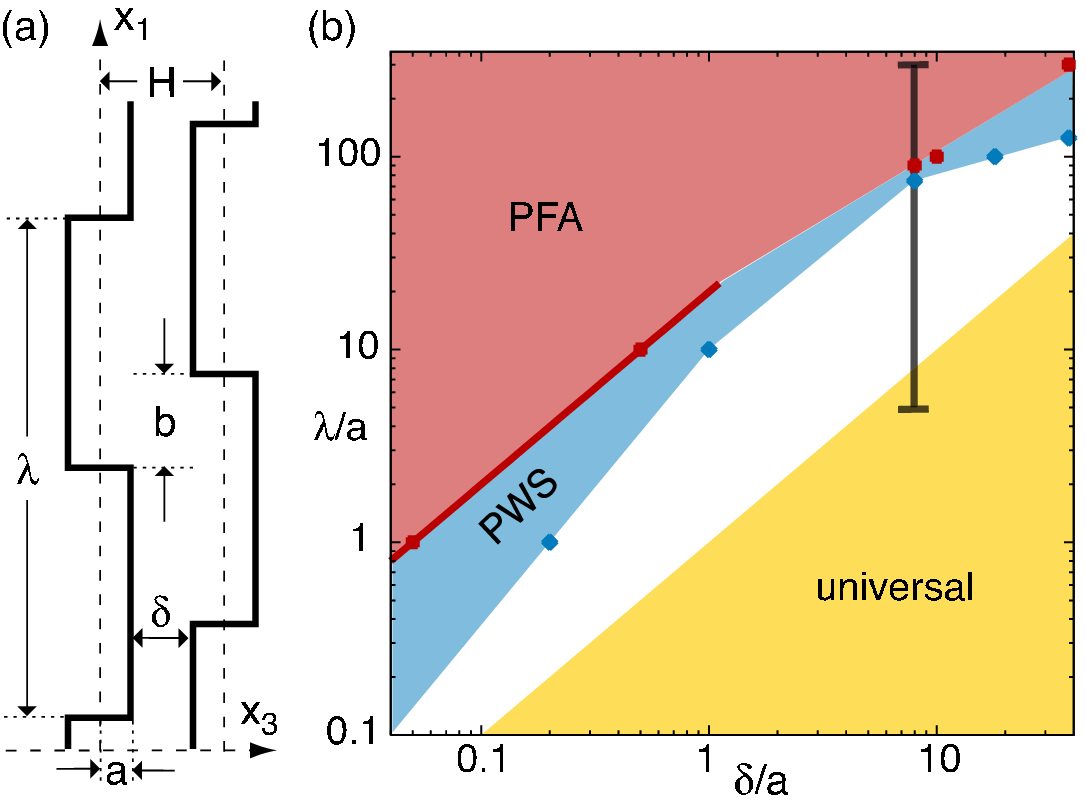}
\includegraphics[width=0.31\linewidth]{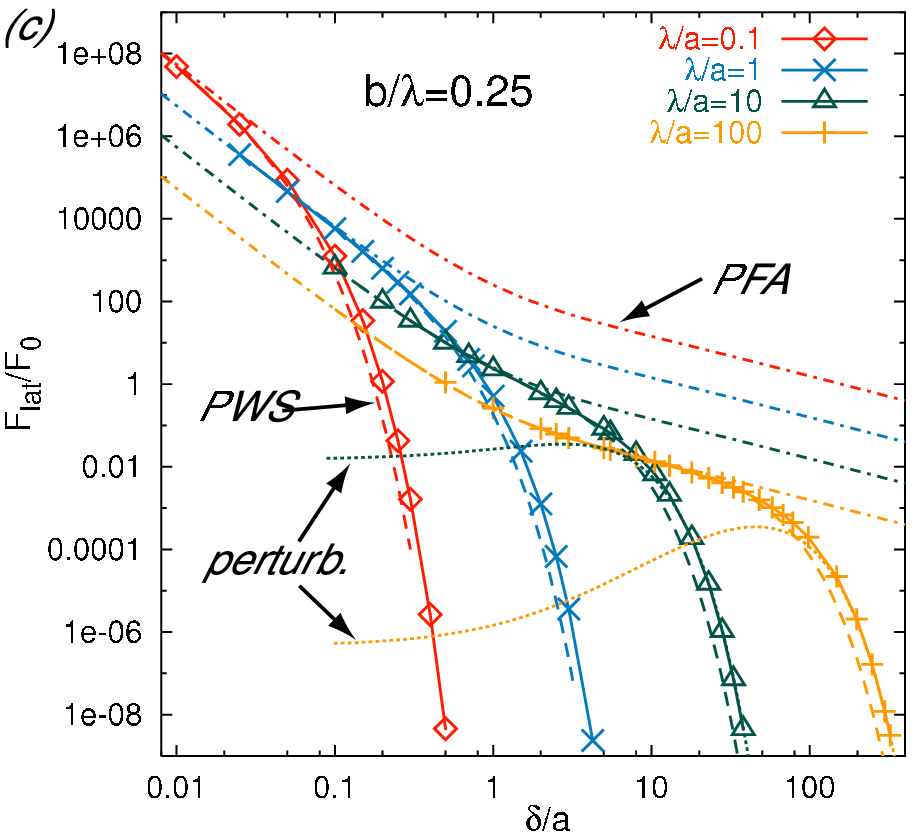}
\includegraphics[width=0.31\linewidth]{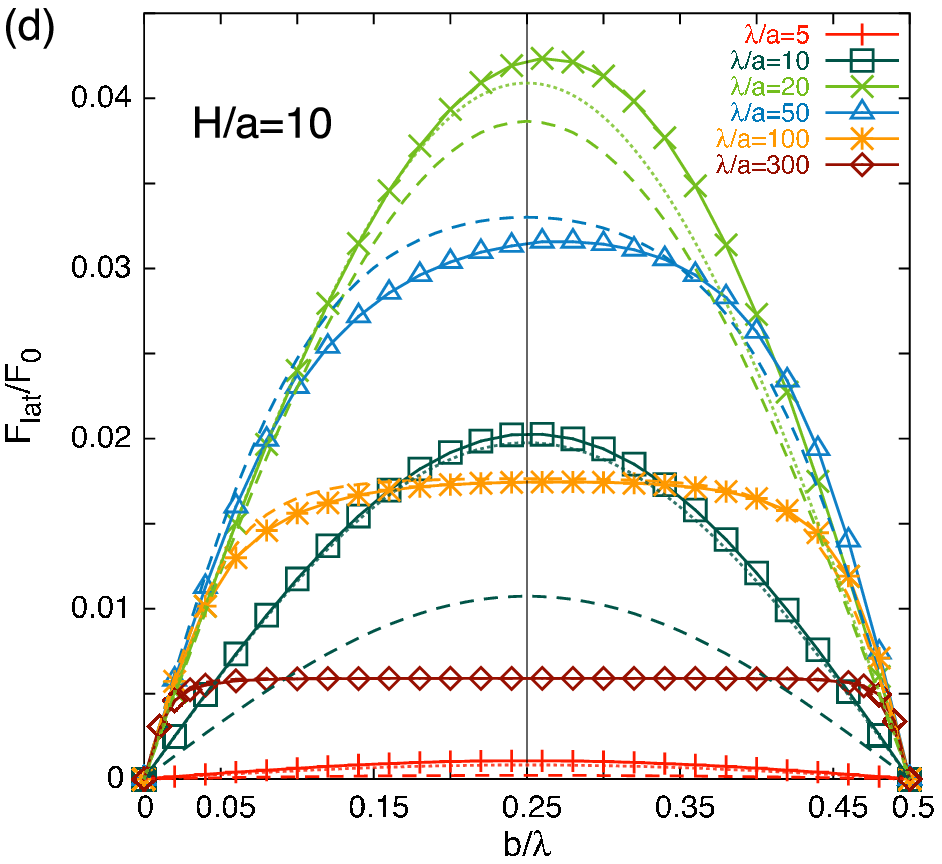}
\caption{\label{fig:fig3}(a) Two plate geometry. (b) Approximate
  validity ranges of PFA and PWS, and the universal parameter
  range. The vertical bar marks the parameter range for $\lambda/a$ in
  (d). (c) Lateral force (in units of the normal force $F_0(d)$
  between flat surfaces). Shown are also the PFA (dash-dotted), PWS
  (dashed) and the universal limit $F_{\rm pt}$ (dotted). (d) Shape
  dependence of the lateral force. The dashed and the dotted curves
  are the PWS and the full perturbative result of \cite{Emig+01+03},
  respectively.}
\end{center}
\end{figure}

Finally, we use the previous results to study the Casimir force for
the geometry in Fig.~\ref{fig:fig3}(a) which emerges from the
variation of the zero-point energy with the shift $b$. The energy can
be obtained from the DOS via Eq.~\eqref{eq:general-energy}.  Without
giving further details we note that the lateral force $F_l=-\partial
\Ecal/\partial b$ can be obtained by computing directly the derivative
of the matrices $N_m^{\alpha\beta}$ in Eq.~\eqref{eq:Bloch} with
respect to $b$. The results are shown as a function of the surface
distance for a fixed $b=\lambda/4$ in Fig.~\ref{fig:fig3}(c) and at a
fixed $H=10a$ as a function of $b$ in Fig.~\ref{fig:fig3}(d).  Let us
compare to approximative methods for small $H$.  The proximity force
approximation \cite{Bordag+01} yields the lateral force $F_{\rm
  PFA}=A[2\Ecal_0(H)-\Ecal_0(H-2a)-\Ecal_0(H+2a)]/\lambda$ for
$0<b<\lambda/2$ with $\Ecal_0(H)=-(\pi^2/720)\hbar c/H^3$, which
changes sign at $b=\lambda/2$ discontinuously.  The pairwise summation
(PWS) of Casimir-Polder potentials $U(r)=-(\pi/24)\hbar c/r^7$ is
often naively applied to metals, using an ``corrected'' amplitude as
to reproduce the correct result for flat ideal metal plates
\cite{Bordag+01}. For the present geometry $F_{\rm PWS}$ can be
computed by a simple numerical integration. The two approximations are
also shown in Fig.~\ref{fig:fig3}(c),(d).  For small $\delta$, both
approximations agree and match our results.  Beyond $\delta \gtrsim
\lambda/20$ the PFA starts to fail since it does not capture the
exponential decay of $F_{\rm lat}$ for increasing $\delta$. The PWS
approach has a slightly larger validity range
[cf.~Fig.\ref{fig:fig3}(b)] and reproduces the exponential
decay. However it deviates by at least {\it one order of magnitude}
from $F_{\rm lat}$ for $\delta \gtrsim 2.5 \lambda$.

At large $\delta \gtrsim \lambda$ the lateral force shows universal
behavior in the form of approaching a limiting form which is
independent of the detailed shape of the surface corrugation.
The universal limit is given by the
lateral force for a {\it sinusoidally}
shaped surface (with amplitude $a_0$ and wavelength $\lambda$)
\cite{Emig+01+03},
\begin{equation}
\label{eq:f-pert}
F_{\rm pt}=\frac{8\pi^3\, \hbar c}{15} \frac{a_0^2 A}{\lambda^5 H}\,
\sin\left(\frac{2\pi}{\lambda} \, b\right) e^{-2\pi H/\lambda} \, ,
\end{equation}
which is expected to hold for $a_0\ll \lambda$.  Using the lowest
harmonic of the present rectangular geometry, corresponding to
$a_0=4a/\pi$, we find indeed excellent agreement between $F_{\rm pt}$
and our results for the geometry of Fig.\ref{fig:fig3}(a) for large
$\delta \gtrsim \lambda$, see Fig.\ref{fig:fig3}(c).

For the dependence on $b$, see Fig.~\ref{fig:fig3}(d), three regimes
can be identified. For $\lambda \gg H$, the force profile resembles
almost the rectangular shape of the surfaces, and the PWS approach
yields consistent results. For decreasing $\lambda$, yet larger than
$H$, the force profile becomes asymmetric with respect to
$b=\lambda/4$ and more peaked, signaling the crossover to the
universal regime for $\lambda \lesssim H$ where the force profile
becomes sinusoidal. We note that the PWS approach fails to predict the
asymmetry of the force profile, and the PFA even predicts no variation
for $0<b<\lambda/2$.

\ack 
Support through the Emmy-Noether Program of the DFG under grant
No. EM70/2 is acknowledged.

\end{document}